\documentclass[prb,twocolumn, superscriptaddress, showpacs]{revtex4}
\usepackage{graphicx}
\begin{document}
\title{Ferromagnetism in diluted magnetic semiconductor quantum dot
arrays embedded in semiconductors}
\author{Pin Lyu}
\affiliation{Department of Physics and Institute of Physics and
Applied Physics, Yonsei University, Seoul 120-749, Korea}
\affiliation{Department of Physics and Center for Theoretical
Physics, Jilin University, Changchun 130023, China}
\author{Kyungsun Moon}
\affiliation{Department of Physics and Institute of Physics and
Applied Physics, Yonsei University, Seoul 120-749, Korea}
\date{October 29, 2002}
\begin{abstract}
We present an Anderson-type model Hamiltonian with exchange
coupling between the localized spins and the confined holes in the
quantum dots to study the ferromagnetism in diluted magnetic
semiconductor (DMS) quantum dot arrays embedded in semiconductors.
The hybridization between the quantum-confined holes in the DMS
quantum dots and the itinerant holes in the semiconductor valence
band makes hole transfer between quantum dots, which can induce
the long range ferromagnetic order of the localized spins. In
addition, it makes the carrier spins both in the DMS quantum dots
and in the semiconductors polarized. The spontaneous magnetization
of the localized spins and the spin polarization of the holes are
calculated using both the Weiss mean field approximation and the
self-consistent spin wave approximation, which are developed for
the present model.

\end{abstract}
\maketitle

\section{introduction}
Diluted magnetic semiconductors (DMS) and a variety of quantum
nanostructures based on DMS materials have recently attracted much
interest due to the novel physics and the potential application to
the emergent field of spintronics.\cite{wolf} The discovery of the
ferromagnetism in the III-V based DMS materials
Ga$_{1-x}$Mn$_{x}$As and In$_{1-x}$Mn$_{x}$As has made it possible
to combine the magnetic and semiconducting properties in one
material. \cite{ohno}  At high concentration of randomly
distributed ${\rm Mn}^{2+}$ ions doped in GaAs samples with high
hole density, Ga$_{1-x}$Mn$_{x}$As compounds exhibit
ferromagnetism with transition temperature as high as 110 K at
certain value of $x$. The ferromagnetism in the localized spins of
impurity Mn ions is mediated by the itinerant holes through the
$p$-$d$ exchange coupling between the valance-band holes and the
localized spins.\cite{ohno,4,Mac,Bhatt} Bhatt \textit{et al.}
argues that for low carrier density and strong disorder limit, the
ferromagnetism is mediated by the carriers in impurity
bands.\cite{Bhatt} Recently the room temperature ferromagnetism
was reported in the Mn-doped magnetic semiconductors
(Ga,Mn)N,\cite{apl} which also belong to the III-V family.

As one of the interesting quantum structures made of the DMS
materials, the self-organized (In,Mn)As quantum dots were
successfully fabricated by growing (In,Mn)As on the (100), (211)B,
(311)B GaAs substrates using low-temperature molecular beam
epitaxy, which were subsequently capped by the GaAs
layer.\cite{guo1} The electron diffraction pattern and the atomic
force microscopy measurement confirmed the formation of the
(In,Mn)As quantum dots. The more detailed experiments afterwards
revealed that the majority of the Mn atoms were actually
substituted in the In-site in (In,Mn)As quantum dots shown by the
fluorescence extended x-ray absorption fine structure
analysis.\cite{guo2} In their samples,\cite{guo1, guo2} the
(In,Mn)As quantum dots were irregularly placed and embedded in the
GaAs. The major challenge in experiment is to fabricate the
uniform and regular array of DMS quantum dots and to explore the
various physical properties of these systems. The interplay
between the quantum-confined magnetic dots and the non-magnetic
semiconductors can be very interesting, which has a possible
application as a tool to implement quantum bits, a large-scale
quantum computer,\cite{loss1,loss2} and other quantum devices.

In this paper, we theoretically study the ferromagnetism in
diluted magnetic semiconductor quantum dot arrays embedded in
semiconductors. In contrast to the bulk ferromagnetic
semiconductors, the following two processes lead to the
ferromagnetism in the present systems: (1) the localized spins of
Mn$^{2+}$ interact with the quantum-confined holes in the dots
through exchange coupling, and (2) the localized orbital of the
quantum-confined holes in the dots hybridizes with the itinerant
holes in the semiconductor valence band. The hybridization allows
hole transfer between the DMS quantum dots, which may induce the
long range ferromagnetic order of the localized spins. In turn,
the carrier spins both in the DMS quantum dots and in the
semiconductors are polarized, which is crucial to realize
spintronics. In order to describe the basic physics, we propose an
Anderson-like model Hamiltonian, where the onsite Coulomb
interaction is replaced by the exchange interaction. Our results
provide a basis for the experiments on exploring the ferromagnetic
properties of these systems.

The paper is organized as follows. In Sec.\ II, an Anderson-type
model Hamiltonian is introduced to describe the DMS quantum dot
arrays embedded in semiconductors. We use the Weiss mean field
approximation and the self-consistent spin wave approximation,
which are modified for the present model. In Sec.\ III, we show
the numerical results of the temperature dependence of various
physical quantities: the spontaneous magnetization of the
localized spins and the spin polarizations of the carriers both in
the quantum dots and in the semiconductor. Finally we conclude
with a brief summary in Sec.\ IV.

\section{Theoretical model and self-consistent approximation}

We consider the regular arrays of DMS quantum dots with the simple
cubic structure embedded in the semiconductor such as (Ga,Mn)As or
(In,Mn)As quantum dots in the GaAs Layer. The model Hamiltonian of
the system can be written by
\begin{eqnarray}
H&=&\sum _{\bf k \sigma} (\varepsilon_{\bf k}-\mu)
c^{\dagger}_{\bf k \sigma}c_{\bf k \sigma}
\nonumber\\
&&+ \sum_{i\sigma}
(\varepsilon_{d}-\mu)d^{\dagger}_{i\sigma}d_{i\sigma}+J\sum _{i}
{\bf S}_{i}\cdot {\bf s}_{i}
\nonumber \\
&&+\sum _{i j\sigma}V\delta ({\bf R}_{i}-{\bf
r}_{j})(c^{\dagger}_{j \sigma}d_{i\sigma}+{\rm H.c.}).
\label{hamiltonian}
\end{eqnarray}
Here $c_{{\bf k}\sigma}$ and $d_{i \sigma}$ are the fermion
operators for the carriers in the semiconductor and in the DMS
quantum dots, respectively. For simplicity, we use the parabolic
band for the carriers in the semiconductor $\varepsilon_{\bf
k}=\hbar {\bf k}^2/(2m^*)$ with the effective mass $m^{*}$.
$\varepsilon_{d}$ is the discrete energy level of the carriers
within the quantum dots,  and $\mu$ is the chemical potential. The
exchange coupling between the confined holes and
${\text{Mn}}^{2+}$ ion impurity spins is special for the DMS
quantum dot system and $J$ stands for the exchange coupling
strength. ${\bf S}_{i}$ is the local spin of Mn ion impurity and
${\bf s}_{i}$ represents the spin of the confined holes at the
{\em i-th} site of quantum dot arrays, which can be written as
$\frac{1}{2}\sum _{\sigma \sigma '} d ^{\dag}_{i \sigma} \vec
{\tau}_{\sigma \sigma '}d_{i \sigma '}$, where $\vec {\tau}$ are
the three Pauli spin matrices. ${\bf R}_{i}$ represents the {\em
i-th} site of the DMS quantum dots in the arrays. The last term in
the Hamiltonian (\ref{hamiltonian}) takes into account the
hybridization between the holes confined in the quantum dots and
the itinerant holes in the semiconductor valence band. The present
Hamiltonian (\ref{hamiltonian}) is quite similar to that of the
periodic Anderson model for the heavy fermion compounds \cite{rmp}
with the onsite screened Coulomb interaction. Here the onsite
interaction is replaced by the exchange coupling between the
localized spin and the hole confined in quantum dots.

Performing the coarse graining procedures \cite {Mac} and using
the Holstein-Primakoff transformations,\cite{auerbach} the
localized spin ${\bf S}_{i}$ can be written in terms of the
bosonic operators $a^{\dag}_{i}, a_{i}$ as follows
\begin{equation}
S^{z}_{i}=cS-a^{\dag}_{i}a_{i},
\end{equation}
\begin{equation}S^{+}_{i}=\sqrt{2cS-a^{\dag}_{i}a_{i}}~a_{i},
\end{equation}
\begin{equation}
S^{-}_{i}= a^{\dag}_{i} \sqrt{2cS-a^{\dag}_{i}a_{i}},
\end{equation}
where $c$ is the mean number of the magnetic ions Mn$^{2+}$ in the
DMS quantum dots. In the following, the approximation
$\sqrt{2cS-a^{\dag}_{i}a_{i}}\simeq \sqrt{2cS}$ is applied.

By transforming from the lattice space to the momentum space and
limiting the involved momentum of the carriers in the
semiconductor valence band within the first Brillouin zone of the
corresponding quantum dot arrays, the Hamiltonian can be rewritten
as follows,
\begin{eqnarray}
H&=&\sum _{\bf k \sigma} (\varepsilon_{\bf k}-\mu)
c^{\dagger}_{\bf k \sigma}c_{\bf k \sigma} + \sum _{{\bf k}
\sigma} (\varepsilon_{d \sigma }-\mu)d^{\dagger}_{{\bf
k}\sigma}d_{{\bf k}\sigma}
\nonumber \\
&&+\sum _{{\bf k}\sigma}V_{\text{eff}}(c^{\dagger}_{{\bf
k}\sigma}d_{{\bf k}\sigma}+{\rm H.c.})
\nonumber \\
&&+\frac{J}{2}\frac{\sqrt{2cS}}{\sqrt{M}}\sum _{\bf qk}
(a^{\dagger}_{\bf q} d^{\dagger}_{{\bf k}\uparrow}d_{{\bf k+q}
\downarrow} +a_{\bf q}d^{\dagger}_{{\bf k+q}\downarrow} d_{{\bf
k}\uparrow})
\nonumber \\
&& -\frac{J}{2M}\sum_{{\bf kq}_{1}{\bf q}_{2}\sigma}\sigma
a^{\dagger} _{{\bf q}_{1}}a_{{\bf q}_{2}}d^{\dagger}_{{\bf
k-q}_{1} \sigma}d_{{\bf k-q}_{2} \sigma}, \label{hami}
\end{eqnarray}
where $\varepsilon_{d\sigma}=\varepsilon_{d}
+\frac{\Delta}{2}\sigma$ with $\Delta=JcS$,
$V_{\text{eff}}=V\sqrt{a^3/b^3}$, $b$ is the lattice constant of
the quantum dot arrays, $a$ the effective radius of hybridization,
and $M$ the total number of dot sites. The summations of $q$,
$q_1$, and $q_2$ in the last two terms in the Hamiltonian
(\ref{hami}) are restricted to the values less than the Debye
cutoff $q_{c}$ for the spin waves of local impurity spins with the
relation $q_{c}^3=6\pi^2 c/b^3$.

In functional integral representation, the partition function $Z$
for the Hamiltonian (\ref{hami}) is given by
\begin{equation}
Z=\int {\cal D}[c^{\dagger}c]{\cal D}[d^{\dagger}d]{\cal
D}[a^{\dagger}a]e^{-\int ^{\beta}_{0}d\tau
L(c^{\dagger}c,d^{\dagger}d,a^{\dagger}a)},
\end{equation}
where the Lagrangian $L$ can be written as
\begin{eqnarray}
L&=&\sum _{{\bf k} \sigma}(c_{{\bf k}\sigma}^{\dagger}\partial
_{\tau}c_{{\bf k}\sigma}+d_{{\bf k}\sigma}^{\dagger}\partial
_{\tau}d_{{\bf k}\sigma})+\sum _{\bf q}a_{\bf q}^{\dagger}\partial
_{\tau}a_{\bf q}
\nonumber \\
&& +H(c^{\dagger}c,d^{\dagger}d,a^{\dagger}a).
\end{eqnarray}
Here the fermionic and bosonic degrees of freedom are represented
by the Grassmann variables and the complex variables,
respectively.

The itinerant carrier degrees of freedom in the semiconductor
valence band can be integrated out easily, and we have
\begin{equation}
Z=Z_{s}\int {\cal D}[d^{\dagger}d]{\cal D}[a^{\dagger}a]e^{-\int
^{\beta}_{0}d\tau L(d^{\dagger}d,a^{\dagger}a)},
\end{equation}
where $Z_{s}$ is the partition function for the free carriers in
the semiconductor valence band, and $L[d^{\dagger}d,a^{\dagger}a]$
is given by
\begin{eqnarray}
L&=&\sum _{{\bf k} \sigma}d_{{\bf k}\sigma}^{\dagger}[
\partial _{\tau}-\mu + \varepsilon_{d \sigma}-V_{\text{eff}}^{2}(\partial
_{\tau}-\mu+ \varepsilon_{{\bf k}})^{-1}] d_{{\bf k}\sigma}
\nonumber \\
&&+\sum _{\bf q}a_{\bf q}^{\dagger}\partial _{\tau}a_{\bf q}
+\frac{J }{2}\frac{\sqrt{2cS}}{\sqrt{M}}\sum _{\bf qk}
(a^{\dagger}_{\bf q} d^{\dagger}_{{\bf k}\uparrow}d_{{\bf k+q}
\downarrow}
\nonumber \\
&& +a_{\bf q}d^{\dagger}_{{\bf k+q}\downarrow} d_{{\bf
k}\uparrow}) -\frac{J}{2M}\sum_{{\bf kq}_{1}{\bf
q}_{2}\sigma}\sigma a^{\dagger} _{{\bf q}_{1}}a_{{\bf
q}_{2}}d^{\dagger}_{{\bf k-q}_{1} \sigma}d_{{\bf k-q}_{2}
\sigma},~~~~
\end{eqnarray}
where $(\partial _{\tau}-\mu+ \varepsilon_{{\bf k} \sigma})^{-1}$
is the Green's function of the itinerant carriers in the
semiconductor valence band.

By subsequently integrating out the remaining carrier degrees of
freedom in the DMS quantum dots, we finally obtain the following
partition function
\begin{equation}
Z=Z_{s}\int {\cal D}[a^{\dagger}a]e^{-S_{\rm eff}},
\end{equation}
where the effective action is given by
\begin{equation}
S_{\text{eff}}=\int _{0}^{\beta} d\tau \sum _{\bf q}
a^{\dagger}_{\bf q}\partial _{\tau} a_{\bf q} -{\rm Tr}{\rm
ln}(G^{d})^{-1}-{\rm Tr}{\rm ln}[(1+G^{d}\delta G^{-1}].
\label{expanding}
\end{equation}
The mean-field part $(G^{d})^{-1}$ is written by
\begin{equation}
(G^{d})^{-1}=[
\partial _{\tau}-\mu + \varepsilon_{d}-V^{2}_{\text{eff}}(\partial
_{\tau}-\mu+ \varepsilon_{{\bf k}})^{-1}] {\bf 1}+\frac{\Delta}{2}
\tau ^{z},
\end{equation}
and the fluctuation part $\delta G^{-1}$ is given by
\begin{eqnarray}
\langle {\bf k}|\delta G^{-1}|{\bf
k'}\rangle&=&\frac{J}{2}\frac{\sqrt{2cS}}{\sqrt{M}}\left
(a^{\dagger}_{\bf k'-k}\frac{\tau ^{+} }{2}+a_{\bf k-k'}\frac{\tau
^{-}}{2}\right)
\nonumber \\
&& - \frac{J}{2M}\sum _{\bf q}a^{\dagger}_{{\bf q}-{\bf k}}a_{{\bf
q}-{\bf k'}}\tau ^{z}.
\end{eqnarray}
Hence the exchange coupling and the hybridization of the carriers
induce the effective non-local interaction between the magnetic
impurities.

Expanding Eq.\ (\ref{expanding}) to the quadratic order in
$a^{\dag}$ and $a$, we obtain
\begin{equation} S_{\text{eff}}=\frac{1}{\beta}\sum _{{\bf q}i\nu _{m}}
a^{\dagger}_{\bf q}(i\nu _{m} )D^{-1}({\bf q},i\nu _{m})a_{\bf
q}(i\nu _{m}),
\end{equation}
where the inverse of the spin wave propagator is given by
\begin{eqnarray}
D^{-1}({\bf q},i\nu _{m})&=&-i\nu _{m}+\frac{J}{2\beta M}\sum
_{{\bf k}\sigma i\omega _{n}}\sigma G^{d}_{\sigma}({\bf k},i\omega
_{n})e^{i\omega_{n}0^{+}}
\nonumber \\
&& +\frac{J \Delta}{2\beta M}\sum _{{\bf k} i\omega
_{n}}G^{d}_{\uparrow}({\bf k},i\omega _{n})
\nonumber \\
&&\times G^{d}_{\downarrow}({\bf k+q},i\omega _{n}+i\nu _{m}),
\end{eqnarray}
and the Green function $G^{d}_{\sigma}({\bf k}, i\omega _{n})$ is
written by
\begin{equation}
G^{d}_{\sigma}({\bf k}, i\omega _{n})=\frac{-1}{i\omega
_{n}-(\varepsilon_{d \sigma}-\mu )-V^{2}_{\text{eff}}/[i\omega
_{n}-(\varepsilon_{\bf k}-\mu)]}.
\end{equation}

The spin wave dispersion is obtained by the following analytic
continuation $i\nu _{m}\rightarrow \Omega +i0^{+}$, which is given
by
\begin{eqnarray}
\Omega _{\bf q}&= &\frac{J}{2\beta M}\sum _{{\bf k} \sigma i\omega
_{n}}\sigma G^{d}_{\sigma}({\bf k},i\omega
_{n})e^{i\omega_{n}0^{+}} +\frac{J \Delta}{2\beta M}
\nonumber \\
&& \times \sum _{{\bf k} i\omega _{n}}G^{d}_{\uparrow}({\bf
k},i\omega _{n}) G^{d}_{\downarrow}({\bf k+q},i\omega _{n}+\Omega
_{\bf q}).~~~~~~\label{spin}
\end{eqnarray}
After the Matsubara frequency summation, the spin wave dispersion
can be derived as follows
\begin{eqnarray}
\Omega _{\bf q}&=&-\frac{J}{2M}\sum _{{\bf k}\sigma \alpha} \sigma
A^{\alpha}_{{\bf k}\sigma}f(\varepsilon ^{\alpha}_{{\bf k}
\sigma}) -\frac{J\Delta \lambda ({\bf q})}{2 M}
\nonumber \\
&& \times \sum _{{\bf k}\alpha} \Biggl [
 A^{\alpha}_{{\bf k}\uparrow}A^{\alpha}_{{\bf k+q}\downarrow}
 \frac{f(\varepsilon ^{\alpha}_{{\bf k}+{\bf
q}\downarrow})-f(\varepsilon ^{\alpha}_{{\bf k}\uparrow})}{\Omega
_{\bf q}+\varepsilon ^{\alpha}_{{\bf k}\uparrow}-\varepsilon
^{\alpha}_{{\bf k}+{\bf q}\downarrow}}
\nonumber \\
&&+A^{\alpha}_{{\bf k}\uparrow}A^{-\alpha}_{{\bf k+q}\downarrow}
\frac{f(\varepsilon ^{-\alpha}_{{\bf k}+{\bf
q}\downarrow})-f(\varepsilon ^{\alpha}_{{\bf k}\uparrow})}{\Omega
_{\bf q}+\varepsilon ^{\alpha}_{{\bf k}\uparrow}-\varepsilon
^{-\alpha}_{{\bf k}+{\bf q}\downarrow}} \Biggl ], \label{spinwave}
\end{eqnarray}
with $A^{+}_{{\bf k}\sigma}$ and $A^{-}_{{\bf k}\sigma}$ are given
by
\begin{equation}
A^{+}_{{\bf k}\sigma}=\frac{\varepsilon ^{+} _{{\bf
k}\sigma}-\varepsilon _{\bf k}+\mu}{\varepsilon ^{+}_{{\bf
k}\sigma}-\varepsilon ^{-}_{{\bf k}\sigma}},
\end{equation}
and
\begin{equation}
A^{-}_{{\bf k}\sigma}=\frac{\varepsilon _{\bf k}-\mu-\varepsilon
^{-} _{{\bf k}\sigma}}{\varepsilon ^{+}_{{\bf
k}\sigma}-\varepsilon ^{-}_{{\bf k}\sigma}},
\end{equation} where the lower band energy is $\varepsilon
^{-}_{{\bf k}\sigma}=\frac{1}{2}(\varepsilon
_{d\sigma}+\varepsilon _{\bf k})-\frac{1}{2}\sqrt{(\varepsilon _{d
\sigma}-\varepsilon _{\bf k})^2+4V^2_{\text{eff}}}-\mu$ and the
upper band $\varepsilon ^{+}_{{\bf
k}\sigma}=\frac{1}{2}(\varepsilon _{d\sigma}+\varepsilon _{\bf
k})+\frac{1}{2}\sqrt{(\varepsilon _{d \sigma}-\varepsilon _{\bf
k})^2+4V^2_{\text{eff}}}-\mu$. The index $\alpha$ represents $+$
or $-$ for upper and lower hybridization bands, respectively and
$f(x)$ is the Fermi-Dirac distribution function. If the last
summation terms in Eq.\ (\ref{spinwave}) are ignored, we obtain
the Weiss mean field approximation with the Weiss mean field given
by $\Omega =-J/(2M)\sum _{{\bf k}\sigma \alpha} \sigma
A^{\alpha}_{{\bf k}\sigma}f(\varepsilon ^{\alpha}_{{\bf k}
\sigma})$, which is proportional to the spin polarization of the
carriers in the DMS quantum dots. Here $\lambda ({\bf q})$ is
introduced as a phenomenological renormalization factor in the
last summation terms in Eq.\ (\ref{spinwave}), which effectively
takes into account the higher order corrections in the
fluctuations in $S^{x}$ and $S^{y}$: $\lambda ({\bf q})  =
\lambda_0 +(1-\lambda_0)\tanh^2 (q/q_c )$. Since the Hamiltonian
of the system is spin rotationally invariant and the ground state
possesses a spontaneously broken symmetry, the system has a
gapless Goldstone mode.\cite{auerbach} The parameter $\lambda_0$
can be determined by imposing the following condition that
$\Omega_{\bf q} = 0 $ as $\bf q$ goes to zero. At large values of
$\bf q$, the renormalizations in the fluctuation term are
negligible and so $\lambda ({\bf q})$ approaches to one. It
appeared that for bulk ferromagnetic semiconductors, no gap arises
in the self-consistent spin wave approximation, which we believe
is crucial to produce quantitatively reliable results.\cite{Mac}
Hence the modification of the above approximation by putting in
the renormlization factor $\lambda ({\bf q})$ can be very
important for the present system.

At finite temperatures, the spin wave dispersion can be
generalized by imposing the following self-consistency conditions
for the finite temperature exchange gap $\Delta (T)$ given by
$\Delta (T)=J_{pd}\langle S^{z} \rangle$. Here $\langle S^{z}
\rangle$ represents the thermal average of the Mn ion spins in the
DMS quantum dot arrays, which are approximately calculated by the
following formula\cite{Mac}
\begin{eqnarray}
\langle{S^{z}}\rangle &=&\frac{1}{M}\sum _{|{\bf q}|<q_{c}}S{\cal
B}_s (\beta S\Omega _{\bf q}) \hspace{2.cm}
\nonumber \\
&=&\frac{1}{M}\sum _{|{\bf q}|<q_{c}}\bigl \{S-n_B (\Omega _{\bf
q})
\nonumber \\
&&+(2S+1)n_B [(2S+1)\Omega _{\bf q}]\bigl \}, \label{aa}
\end{eqnarray}
where ${\cal B}_s (x)$ is the Brillouin function and $n_B (x)$ the
Bose-Einstein distribution function. The second term in the second
equality of Eq.\ (\ref{aa}) describes how the thermally induced
spin waves from the independent Bose statistics with no limit in
the number of spin waves reduce the magnetization of the system.
The third term takes into account the correct spin kinematics,
which rules out the unphysical states. When $\Omega _{\bf q}$ is
independent of ${\bf q}$, for exmaple, the Weiss mean field
approximation, Eq.\ (\ref{aa}) is reduced to the widely used
formula $\langle {S^{z}}\rangle =cS{\cal B}_{s} (\beta S\Omega)$.

On the other hand, using the Hamiltonian (\ref{hami}) without the
exchange coupling, one can integrate out the carrier degrees of
freedom in the DMS quantum dots to obtain the Green's function for
the semiconductor carriers, where the role of the hybridization is
taken into account. It is given by
\begin{equation}
G^{s}_{\sigma}({\bf k}, i\omega _{n})=\frac{-1}{i\omega
_{n}-(\varepsilon_{\bf k}-\mu )-V^{2}_{\text{eff}}/[i\omega
_{n}-(\varepsilon_{d\sigma}-\mu)]}.
\end{equation}
The densities of the carriers in the semiconductors and in the DMS
quantum dots can be calculated from
\begin{equation}
n^{s}_{\sigma}=\frac{1}{\pi {\cal V}} \sum _{\bf k} \int
^{\infty}_{- \infty}d \omega f(\omega) {\rm Im}G^{s}_{\sigma}({\bf
k},\omega+i0^{+}),
\end{equation}

\begin{equation} n^{d}_{\sigma}=\frac{1}{\pi {\cal V}} \sum _{\bf k}\int ^{\infty}_{-
\infty}d \omega f(\omega) {\rm Im}G^{d}_{\sigma}({\bf k
},\omega+i0^{+}),
\end{equation}
where ${\cal V}$ is the volume of the system. The total carrier
density is given by $n_{\rm tot}=\sum _{\sigma}
(n^{s}_{\sigma}+n^{d}_{\sigma})$, which is fixed and determines
the Fermi energy.

The spin polarization of the carriers is defined as
$P_{s}=(n^{s}_{\downarrow}-n^{s}_{\uparrow})/
(n^{s}_{\downarrow}+n^{s}_{\uparrow})$ in the semiconductor and
$P_{d}=(n^{d}_{\downarrow}-n^{d}_{\uparrow})/
(n^{d}_{\downarrow}+n^{d}_{\uparrow})$ in the DMS quantum dots.
This definition of spin polarizations is different from those
commonly used in the transport properties, which is defined as the
corresponding density of states at the Fermi energy. \cite{lyu}

Based on the above framework, one can calculate the spontaneous
magnetization of the localized spins in the DMS quantum dot arrays
and the spin polarizations of the carriers both in the DMS quantum
dots and in the semiconductor by solving the set of coupled
equations in Weiss mean field approximation and self-consistent
spin wave approximations as well.

\section{Results and Discussions}
\begin{figure}
\centerline{\includegraphics[width=6.0cm]{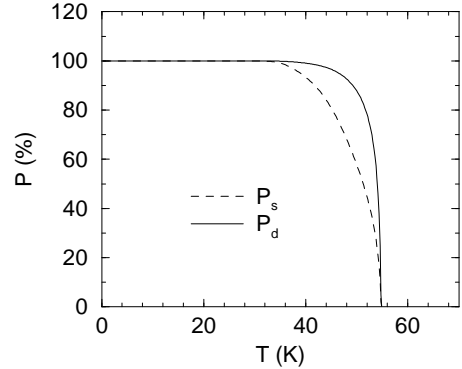}} \caption{Spin
polarization of the carriers in DMS quantum dots ($P_{d}$) and in
semiconductors ($P_{s}$) in Weiss mean field theory with the
following parameters $J=0.15~{\text{eV}}$, $m^{*}=0.5m_{e}$,
$c^{*}=0.1~ {\text{nm}}^{-3}$, $c=1.0$, $b=1.0~{\text{nm}}$,
$V_{\text{eff}}=0.30 ~{\text{eV}}$, and
$\varepsilon_{d}=0.46~{\text{eV}}$.} \label{fig1}
\end{figure}

\begin{figure}
\centerline{\includegraphics[width=6.0cm]{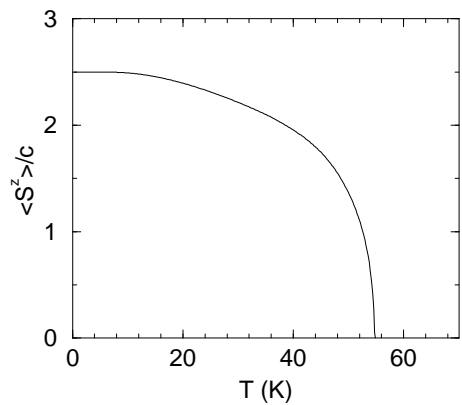}}
\caption{Spontaneous magnetization of the localized spins in DMS
quantum dot arrays in Weiss mean field theory with the same
parameters used in Fig.\ 1.} \label{fig2}
\end{figure}

We have chosen the typical material parameters \cite{ohno} of the
bulk (Ga,Mn)As for the DMS quantum dots, i.e. $J=0.15~ \text{eV}$,
$m^{*}=0.5m_{e}$. The lattice distance $b$ of the DMS quantum dots
is set to be $b=1.0~\text{nm}$, which is within the spin-coherence
length in semiconductor GaAs.\cite{nature} The carrier bands
consist of the upper $\varepsilon ^{+}_{{\bf k}\sigma}$ and lower
$\varepsilon ^{-}_{{\bf k}\sigma}$ bands due to the hybridization
between the discrete energy level of DMS quantum dot and the
semiconductor valence band. The spin dependence of the upper and
lower bands is originated from the exchange coupling in the DMS
quantum dots. Also there exists a gap between the spin-up
(spin-down) upper and lower bands. When the ground state is
metallic or semiconducting, the Fermi level must lie in the
spin-down lower band.

\begin{figure}
\centerline{\includegraphics[width=6.0cm]{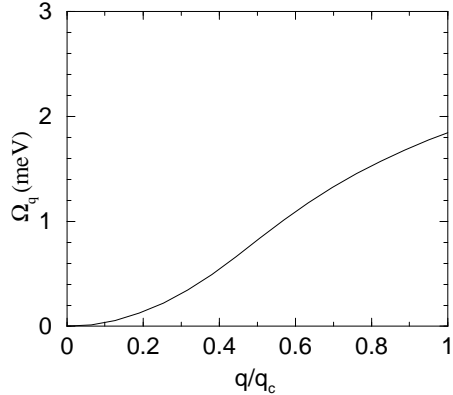}}
\caption{Renormalized spin wave dispersion at 4.2 K in the
self-consistent spin wave approximation with the same parameters
used in Fig.\ 1.} \label{fig3}
\end{figure}

In the following we calculate the ferromagnetic properties of the
present system both in the Weiss mean field approximation and in
the self-consistent spin wave approximation, respectively.

In Fig.\ 1 and Fig.\ 2, we plot the temperature dependence of the
spin polarizations of the carriers in the semiconductor and in the
DMS quantum dots, and the spontaneous magnetization for the DMS
quantum dot arrays in the Weiss mean field theory with the
following material parameters $c^{*}=0.1~ \text{nm}^{-3}$,
 $c=1.0$, $V_{\text{eff}}=0.30
~\text{eV}$, and $\varepsilon_{d}=0.46~\text{eV}$. The critical
temperature $T_c$ for spin polarization is about 55 K. One can
notice that the spin polarizations $P_d$ and $P_s$ decrease more
stiffly than the spontaneous magnetization $\langle{S^{z}}\rangle$
of localized spins. The $T_c$ presumably increases up to the bulk
value as one tunes the exchange coupling strength $J$ and the
hybridization $V_{\text{eff}}$ to higher values.

\begin{figure}
\centerline{\includegraphics[width=6.0cm]{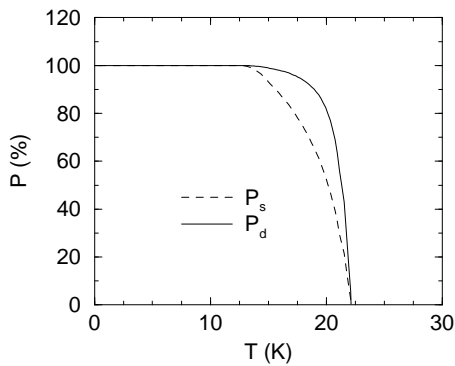}} \caption{
Spin polarization of the carriers in DMS quantum dots ($P_{d}$)
and in semiconductors ($P_{s}$) in the self-consistent spin wave
approximation with the same parameters used in Fig.\ 1.}
\label{fig4}
\end{figure}

\begin{figure}
\centerline{\includegraphics[width=6.0cm]{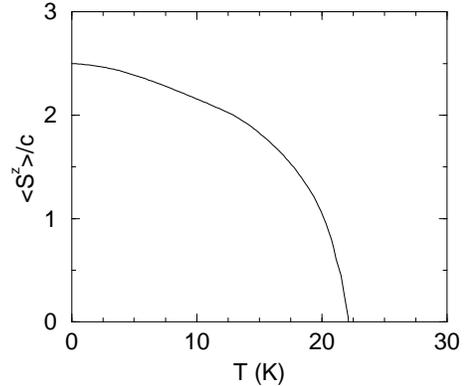}} \caption{
Spontaneous magnetization of the localized spins in DMS quantum
dot arrays in the self-consistent spin wave approximation with the
same parameters used in Fig.\ 1.} \label{fig5}
\end{figure}

\begin{figure}
\centerline{\includegraphics[width=6.0cm]{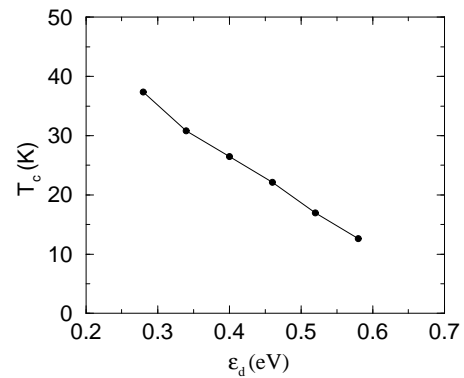}} \caption{
Transition temperature $T_{c}$ versus discrete energy level
$\varepsilon_{d}$ in the self-consistent spin wave approximation
with the following parameters $J=0.15~{\text{eV}}$,
$m^{*}=0.5m_{e}$, $c^{*}=0.1~{\text{nm}}^{-3}$, $c=1.0$,
$b=1.0~{\text{nm}}$, and $V_{\text{eff}}=0.30~{\text{eV}}$.}
\label{fig6}
\end{figure}

Next we go beyond the standard Weiss mean field theory. In Fig.\
3, we have shown the renormalized spin wave dispersion curve at
$T=4.2~\text{K}$ based on the the self-consistent spin wave
approximation. The material parameters are the same as those used
in the Weiss mean field calculation. The spin wave frequencies are
less than the Weiss mean-field value $\Omega \simeq
4.14~{\text{meV}}$. As ${\bf q}$ goes to zero, $\Omega_{\bf q}
\propto q^2$, which are correct for ferromagnetic spin wave
dispersion. As ${\bf q}$ becomes very large, $\Omega_{\bf q}$
approaches to the mean field value. Fig.\ 4 and Fig.\ 5 illustrate
the temperature dependence of the spin polarizations of the
carriers in the semiconductor and in the DMS quantum dots, and the
spontaneous magnetization for the DMS quantum dot arrays in the
self-consistent spin wave approximation with the same material
parameters as before. The magnetization and the spin polarizations
in the self-consistent spin wave approximations drop much rapidly
comparing to those in the Weiss mean field approximation due to
the strong spin fluctuations. The $T_c$ drops to about 22 K less
than a half of the mean field value of $T_c$, which also happens
in the bulk ferromagnetic semiconductors.\cite{Mac} The
qualitative behaviors remain the same as the mean field results.

Fig.\ 6 shows the dependence of the critical temperature $T_c$ as
a function of the discrete energy level $\varepsilon_{d}$ of
quantum dot. The $T_c$ increases with the decrease of
$\varepsilon_{d}$. As $\varepsilon_{d}$ decreases, the carriers in
the semiconductor easily hop to the localized levels and vice
versa, which subsequently enhances the ferromagnetic coupling
between impurity spins. The monotonic dependence of $T_{c}$ on
$\varepsilon_{d}$ is only correct in the low carrier density
region, in which our model is suitable, that is, the Fermi energy
level lying in the spin-down lower band. In the low carrier
density limit, the Fermi energy level always lies below
$\varepsilon_{d \downarrow}$. Obviously the enhancement of the
hybridization strength between the localized carriers in DMS
quantum dots and the itinerant carriers in semiconductor makes
carrier transfer easier leading to the increase of $T_c$.

\section{summary}
We have theoretically studied the origin of the ferromagnetism in
diluted magnetic semiconductor quantum dot arrays embedded in
semiconductors based on an Anderson-type model Hamiltonian. The
hybridization between the quantum-confined holes in the DMS
quantum dots and the itinerant holes in the semiconductor valence
band makes the hole transfer between the DMS quantum dots, which
induces the long range ferromagnetic order of the localized spins
in the DMS quantum dot arrays through the exchange coupling.
Currently the available DMS quantum dot systems have irregular
nanostructures, and the uniform and regular arrays of DMS quantum
dots are expected in experiments. Our results provide a basis for
exploring the magnetic properties of these systems.

\section{acknowledgments}

It is our great pleasure to acknowledge useful discussions with
R.\ N.\ Bhatt and T.\ K.\ Lee. This work was supported by the
Brain Korea 21 Project and Grant No. R01-1999-00018 from the
interdisciplinary Research program of the KOSEF. P.\ L.\ was also
supported by SRF for ROCS from SEM of China and Research
Foundation of Jilin University.

\end{document}